# Quantum information processing in diamond

Jörg Wrachtrup and Fedor Jelezko

*3. Physical Institute, University of Stuttgart, 70550 Stuttgart, Germany*

## Abstract

Quantum computing is an attractive and multidisciplinary field, which became a focus for experimental and theoretical research during last decade. Among other systems, like ions in traps or superconducting circuits, solid-states based qubits are considered to be promising candidates for first experimental tests of quantum hardware. Here we report recent progress in quantum information processing with point defect in diamond. Qubits are defined as single spin states (electron or nuclear). This allows exploring long coherence time (up to seconds for nuclear spins at cryogenic temperatures). In addition, the optical transition between ground and excited electronic states allows coupling of spin degrees of freedom to the state of the electromagnetic field. Such coupling gives access to the spin state readout via spin-selective scattering of photon. This also allows using of spin state as robust memory for flying qubits (photons).

# Introduction

Investigation and characterization of defects has been of uttermost importance for solid state physics and device technology. Often defects determine the mechanical, electrical and optical properties of solids. Advances in material growth and purification techniques have helped to unravel the role of impurities versus bulk material properties. Modern silicon device technology would not have been possible without, for example, the precise control of impurity content. A particular class of defects are optically active defects, so called colour centres. They have been investigated intensively in different materials in the 60' and 70'. The ongoing interest in detailed characterization of colour centres is manly inspired by possible applications in present and future optoelectronic devices (Vavilov, 1994) (GaN based materials are a good example here). Applications of colour clusters for fabrication of laser media is also under discussion (Rand and Deshazer, 1985). If we extend the notion of colour centres to rate earth ions in dielectrics, it is worth mentioning that optical data communication has taken great benefit from colour centres based optical amplifiers. Hence in all areas of modern data processing and communication techniques defects, in one or the other respect, have had substantial impact.

One might ask whether they will play a similar role in the upcoming area of quantum information technology. As a matter of fact, already early proposals (like the Kane  (Kane, 1998) or others solid-state schemes (Shahriar *et al.*, 2002; Stoneham, 2003; Stoneham *et al.*, 2003; Wrachtrup *et al.*, 2001)) are based on defects in solids. They make use of a number of important properties of defects. Due to the localized nature of the electronic wave function associated with defects, dephasing times of optical transitions are usually long and optical resonances are rather narrow. A wide variety of optically active defects is known (especially for wide band-gap semiconductors like diamond or SiC). This allows choosing a system with suitable optical and magnetic properties, such as an electron paramagnetic ground state. A number of defects in dielectric host materials are known with electron spin quantum number S>0. While nuclear spins usually are not directly coupled to optical transitions, photon degrees of freedom might be mapped on to nuclear spin wave functions via hyperfine coupling between electron and nuclear spins. This might be of use in single photon memory and quantum repeater devices. Although important knowledge exists about point defect, only a few of such defects have been investigated in the context of quantum information processing. In this contribution we will focus mainly on colour centres in diamond although other systems do show similar appealing properties.

## NV centers photophysics and spin states

There are more than 100 luminescent defects in diamond (Davies, 1994; Zaitsev, 2001). Many of them have been characterized by optical spectroscopy (Zaitsev, 2000). Quite a number give rise to strong electron spin resonance signals, and for a few even optical detection of magnetic resonance is possible (Pereira *et al.*, 1994). The nitrogen vacancy colour centre is among them (see Figure 1). It has been studied via hole burning (Martin *et al.*, 2000; (Martin, 1999; Yokota *et al.*, 1992) and optical echo spectroscopy (Rand *et al.*, 1994) as well as optically detected magnetic resonance (Vanoort *et al.*, 1990). The bright red fluorescence emitted by the defect with a zero phonon of line at 637 nm is due to an optical transition between spin-triplet states $^3E$ and $^3A$. (Goss *et al.*, 1997). In the absence of a magnetic field the ground symmetry spin state is split by 2.88 GHz into a doublet $X,Y$ ($m_s = \pm 1$) and the third spin sublevel $Z$ ($m_s = 0$)(see Figure 2).

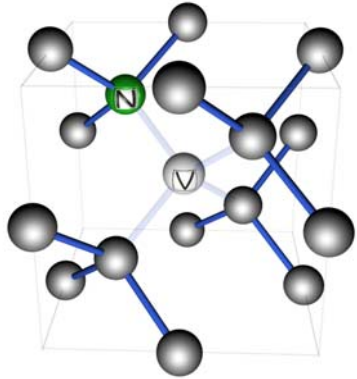

**Figure 1. Structure of nitrogen-vacancy (NV) centre. The single substitutional nitrogen atom (N) is accompanied by a vacancy (V) at a nearest neighbour lattice position.**

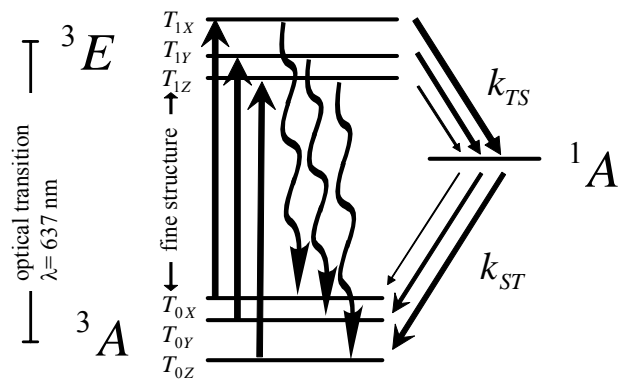

**Figure 2. Energy level diagram of an NV centre. Allowed optical transition between ground ($^3$A ) and excited ($^3$E ) electronic states sublevels are shown. The strength of spin-selective intersystem crossing transitions is encoded in thickness of arrows.**

Optical illumination generates a non-Boltzmann spin alignment of the electron spin in the ground $^3$A state (Harrison *et al.*, 2004). This provides the basis for a number of ODMR experiments including the measurement of the *X,Y,Z* fine structure splitting as well as investigation of electron spin dephasing phenomena (Kennedy *et al.*, 2003; (Charnock and Kennedy, 2001). The origin of the strong spin polarisation has long been subject to intense discussion. Meanwhile it seems to be clear that intersystem crossing to the excited singlet state plays an important role (Harrison *et al.*, 2004; Nizovtsev *et al.*, 2005; (Nizovtsev *et al.*, 2001). Spin-orbit coupling is not playing an essential role in formation of the excited state fine structure (Martin, 1999). Hence an optical dipole transition between $^3A$ and $^3E$ conserves spin angular momentum. The probability for a spin-flip transition i.e. $T_{oi} \rightarrow T_{1g}$ is given by $\langle S_i \, | \, S_g \rangle^2$, which is a measure of the colinearity of the ground and excited state zero-field splitting tensors. There is evidence that both tensors are not perfectly collinear. However, experiments show that conservation of the spin state upon optical excitation is a good assumption (Jelezko *et al.*, 2002; (Jelezko and Wrachtrup, 2004).

For successful modelling of single defect centre experiments it has to be assumed that the intersystem crossing rates of $T_{1X,Y}$ to the singlet 1A state are three orders of magnitude larger than from the $T_{1Z}$ state . In contrast, the relaxation from the $^1A$ state to the triplet ground state is almost isotropic, i.e. rates to the $T_{0i}$ spin sublevels are similar. After a few optical excitation and emission cycles this generates a strong spin polarization (>80%) among the $T_{0X,Y}$ ($m_s\pm1$) and $T_{0Z}$ ($m_s=0$) state. The absorption cross section of the $^3A$-$^3E$ transition at room temperature is around $10^{-16}$ cm$^2$, close to the value for a typical dye molecule. The fluorescence quantum yield is about 70% and a fluorescence lifetime is 13 ns (Collins *et al.*, 1983). These values allow for the observation of single defect centres fluorescence. The fluorescence intensity depends on the spin state because of spin-selective shelving into metastable singlet state. This selectivity is the basis of optical detection of magnetic resonance on single centres (Jelezko and Wrachtrup, 2004). Figure 3 shows the fluorescence of a single defect centre as a function of the applied microwave frequency. When the microwave frequency reaches the resonance with the $m_s=\pm1$ and $m_s=0$, magnetic dipole transitions change the ground-state spin polarization which results in a reduction of the fluorescence intensity. This effect is easily understood by considering that transferring the system from the $m_s=\pm1$ state to the $m_s=0$ state drastically increases the ISC probability. Since the $^1A$ state has a lifetime of some hundred of ns instead of 10 ns for the $^3E$ state (Nizovtsev *et al.*, 2001), a noticeable reduction in the fluorescence intensity is expected. Assuming the above scenario the saturated fluorescence $\left\langle I_{sat}^{flr} \right\rangle$ intensity of a single defect can be written as

$$\left\langle I_{sat}^{flr} \right\rangle = \frac{A}{4 + \dfrac{k_s k_D + R}{R k_T}} \, .$$

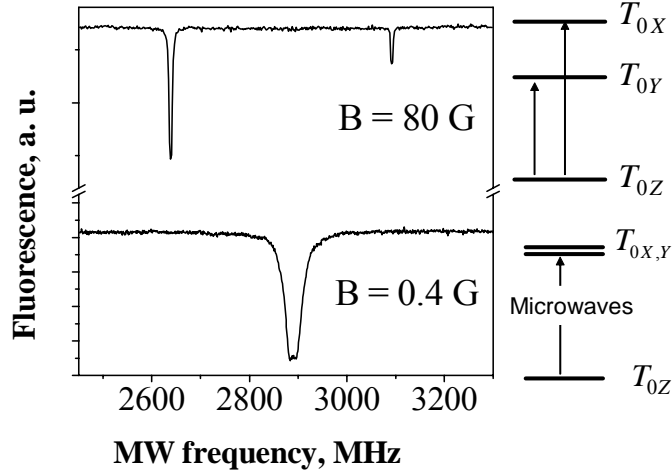

**Figure 3. Optically detected ESR spectra of single NV defects in zero (lower graph) and weak (upper graph) magnetic fields. Relevant spin transitions are shown on the right side of the graph.**

Here, $A$ is the Einstein coefficient, $k_s$ is the ISC rate of the $T_{1z}$ state to the excited singlet state $^1A$, $k_D$ is given by $k_D = k_X + k_Y$, where $k_X$, $k_Y$ and $k_Z$ are the singlet depopulation rates to the $T_{0X,Y,Z}$ states. $R$ is the spin lattice relaxation rate, $k_T = k_D + k_Z$. At low temperature R is expected to be much smaller than $k_s$, $k_D$ and $k_z$ such that

$$\left\langle I_{sat}^{flr} \right\rangle \sim \frac{RA}{k_s}.$$

Substituting typical values for $R \sim 1 s^{-1}$ gives an estimate for the low temperature fluorescence rate. Taking into account that the detection efficiency is on the order of 0,1%, $k_s$ must not exceed a few kHz to allow for detection of the single centre low-temperature fluorescence excitation line. Figure 4 shows an example of a fluorescence excitation line at T=2K. First of all, as predicted by the model, only a single excitation line is visible, although there are three allowed optical transitions ($T_{0j}$ to $T_{1j}$). Even if one assumes that no spin cross terms are allowed $\left\langle T_{oi} \middle| \vec{p} \middle| T_{oi} \right\rangle = 0$ with $\vec{p}$ the electric dipole operator one might detect a multiple line structure because the $D$ and $E$ values in ground and excited state are not exactly equivalent (Martin, 1999).

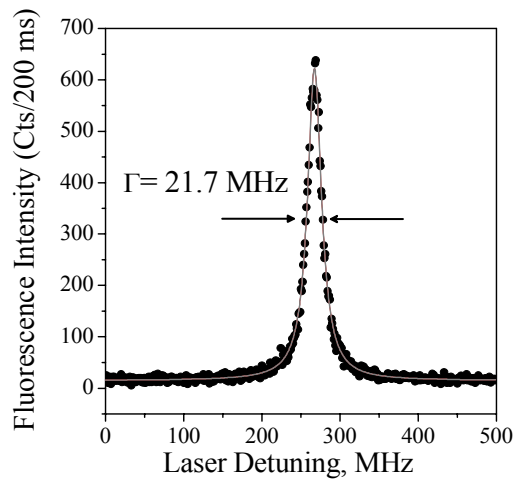

**Figure 4. Fluorescence excitation line of a single NV defect in diamond at T= 1.6 K (circles). Fit curve (solid line) is a Lorentzian function with linewidth (FWHM) of 21.7 MHz.**

The observation of a single excitation line underlines once again the assumption made in the beginning. The ISC rates from the $T_{1Z}$ state, $k_s$, is too large to allow for a detectable $\left\langle I_{sat}^{flr} \right\rangle$ of this state. Since the other two states $T_{1X,Y}$ are degenerated they give rise to a single excitation line observed in the experiment. The resonance linewidth in Figure 4 is around 21 MHz, close to the value expected from the excited state lifetime (~24MHz). It should be noted, that unlike at room temperature, at T=2 K the defect centre emits fluorescence in a telegraph like fashion. When the excitation laser is in resonance with the $T_{0,X,Y}$ and $T_{1,X,Y}$ transition, fluorescence intermittences are observed. These intermittences stem from spin flips and a yet not fully understood spectral diffusion process, which is probably related to charge fluctuations in the surrounding of the defect.

## Use of defects as single photon emitters

The remarkable photostability of defects in diamond makes them good candidates for single photon emitters. Such photons sources, which produce single photon pulses on demand, are important for realization of quantum cryptographic protocols and optical quantum information processing schemes (Gisin *et al.*, 2002). Single quantum systems like ions (Maurer *et al.*, 2004), molecules (Brunel *et al.*, 1998) and quantum dots (Michler *et al.*, 2000) are all good candidate structures because their emission is characterized by antibunching of emitted photons for short interleave times. This antibunching is a characteristic feature of electromagnetic radiation originating from a single quantum mechanical two level system. It is caused by the finite time it takes to build up coherence and finally populate an excited state after the two-level system has decayed to its ground state by photoemission. Like the other systems mentioned above, also the NV-centres shows antibunching in the fluorescence emission (Beveratos *et al.*, 2002; Kurtsiefer *et al.*, 2000). High quality single photon emission, i.e. a photon stream with low contamination from background photons, has been achieved for NV centres in very high quality bulk diamond samples or diamond nanocrystals only. This is because the broad emission bandwidth of the NV centre ranging from 640 to 750 nm does not allow efficient spectral filtering from background radiation. Other defects (like the nickel-related NE8 defect, which structure is presented in Figure 5) showing sharp emission line are more promising candidates for further improvement. The NE8 defect comprises a central Ni atom surrounded by four nitrogen atoms (Nadolinny *et al.*, 1999). Because there is no vacancy related to the defect, this colour centre is characterized by low electron-phonon coupling. Hence the room temperature zero-phonon emission line has a width of only 1.5 nm ; (Gaebel *et al.*, 2004). The emission occur around 800 nm. The relative integral intensity of the zero-phonon line to the entire spectrum (Debye-Waller factor) is 0.7 (see fluorescence emission spectrum presented in Figure 6. Hence, most of the emission occurs at zero-phonon line and efficient spectral filtering is possible. The basic photophysical parameters are close to those of the NV centre, notably the excited state lifetime is around 5 ns and the intersystem crossing rate to the metastable state is below 20 MHz. Since the NE8 defect comprises four nitrogen atoms, it is unlikely that it can be implanted like it is the case for the NV centre. Hence it was an important step to show that the NE8 can be "artificially" created during the growth process of a CVD diamond film (Rabeau *et al.*, 2005). Although incorporated in a microcrystalline environment, the NE8 conserved it advantageous photophysical and spectral features.

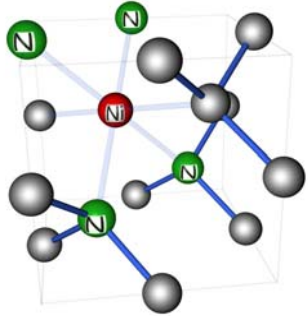

**Figure 5. The structural model of the NE8 nickel–nitrogen defect diamond. The nickel atom is surrounded by four interstitial nitrogen atoms.**

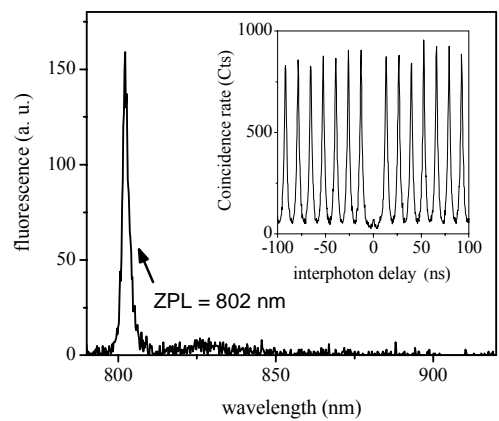

**Figure 6. The fluorescence emission spectrum of NE8 centre recorded at room temperature. Inset shows the fluorescence correlation function recorded using pulsed excitation. The absence of a peak at zero interphoton delay indicates single photon emission.**

## Use of defects as quantum memory and repeaters

Besides the controlled generation of single photons and photon pairs, the development of quantum memories and repeaters will be of crucial importance in future quantum information technology (Fleischhauer and Lukin, 2002; Briegel *et al.*, 1999). While there are promising solutions at hand for single photon emitters, only few developments have been made in the field of memory and repeater devices (Bajcsy *et al.*, 2003). In part this is due to the fact that practical units should work under ambient conditions. Especially solid-state implementation would be preferable because of commercialization. The physics of memory devices is based on a reduction of the group velocity for an incoming photon. The photon phase and amplitude needs to be converted to a spin quantum state, which preserves the information for as long as possible time. Good candidates are rare-earth ions (here storage of quantum state of light for times longer than a second was demonstrated recently (Longdell *et al.*, 2005) and the NV defect centre. No photon to spin state transfer has been shown for the NV centre yet, but initial experiments are in progress and the idea behind them shall be explained below

The most often discussed method for photon-spin conversion is based on electro magnetically induced transparency (EIT). In this scheme an opaque medium is rendered to be transparent by the coherent action of two laser beam coupled to a quantum three level system. Here it should be assumed that such a system is realized by the NV centre by considering two spin ground states and one excited electronic state spin sublevel. EIT in diamond has been shown, both in the microwave (Wei and Manson, 1999) and optical domain (Hemmer *et al.*, 2001). In short, in a three level system where the three levels are coupled by two (near)resonant laser fields, the eigenvalues of the complete Hamiltonian are a symmetric and antisymmetric combination of the two lowest states. While the symmetric combination has a dipole allowed transitions to the excited state, this is not the case for the antisymmetric combination (dark state). Figure 7 shows an example of an EIT-experiment among the electron and nuclear spin sublevels in a single defect. Two MW frequencies in resonance with the *X-Z* transitions have been used. Hyperfine structure sublevels of the *Z* state associated with nitrogen nuclear spin (states $|1\rangle$ and $|3\rangle$) and the *X* state (state $|2\rangle$) form effective lambda-scheme. When the probe field comes close to resonance with the dark level, the two contributions from the coupling and probe fields interfere destructively and an increase in the defect centre fluorescence is seen (see Figure 8). This increased fluorescence is visible as a dip in ODMR signal. The width of the dip should be given by the dephasing time of the nuclear spin, which is on the order of a few seconds. The actual linewidth observed is roughly 1 MHz, mostly due to the large linewidth of the MW sources used.

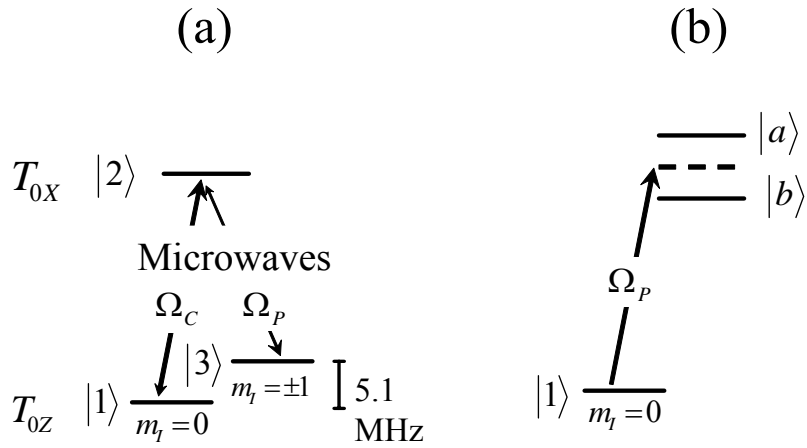

**Figure 7. (a) Partial ground-state energy levels diagram relevant for EIT experiments. Optical pumping was used for polarising the system in Z spin sublevel of the ground electronic state. Microwave coupling field continuously drives transition while the probe filed was swept across resonance. EIT occurs when coupling field-probe field detuning matches the hyperfine structure splitting. (b) EIT experiment in terms of dressed-state representation. For details see text.**

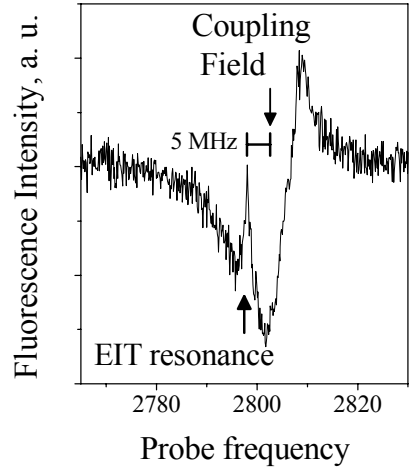

**Figure 8. ESR spectrum for a probe with the coupling field frequency fixed at 2803 MHz. The arrows indicate EIT occurring at 2797 MHz. Derivative lineshape in the high frequency region of the spectrum is related to relatively high Rabi frequency of the probe field**

In the case of EIT, since one of the fields is large it is logical to choose the dressed state basis. In the experiments in Figure 8 the strong field, called the coupling field $\Omega_c$ is in resonance with the $|2\rangle \rightarrow |3\rangle$ transition. In the dressed state picture, the $X$ state forms a coherent superposition, which for resonant coupling, i.e. $\Delta\omega_{x-z} = \omega_c$ is of the form

$$|a\rangle = \frac{1}{\sqrt{2}}|3\rangle + |2\rangle,$$

$$|b\rangle = \frac{1}{\sqrt{2}} |3\rangle - |2\rangle.$$

The transition amplitude at the (undressed) resonant frequency $(|2\rangle - |3\rangle)/\hbar$ from the $|1\rangle$-state to the dressed states will be the sum of contributions of states $|a\rangle$ and $|b\rangle$. Since the $|3\rangle$ level is metastable, the contributions from the $|1\rangle - |2\rangle$ cancel because they enter with opposite signs. This cancellation of absorption on the $|1\rangle - |2\rangle$ transition can be viewed in term of a Fano-type interference.

This demonstration indicates that *EIT-type experiments are indeed feasible on single defect centres* and need to be transferred to the optical domain to be of use in quantum memory applications. In such a scheme the two light field may couple the $m_s = 0$ as well as the $m_s = \pm 1$ in the $^3E$ state. The sharp dip in absorption will create a large (and negative) $\frac{d\omega}{dn}$ and hence slows group velocity $v_{gr} = \frac{c_o}{n(\delta) + w_a \frac{dn}{d\delta}}$. If a pulse at the probe field enters the medium in which all defect centres are initially in the $m_s = 0$ state, the front edge of the pulse will be decelerated. As a result the spatial extend of the pulse will be compressed by the ratio $c/v_{gr}$, where $v_{gr}$ is the group velocity in the medium. The energy in the pulse is much smaller inside the medium since a coherence between the $m_s = 0$ and $m_s = \pm 1$ state needs to be build up. This process is a change of the spin state and energy here is carried away by the control field. The wave of flipped spins inside the medium propagates together with the light pulse. It turns out that such an excitation can be associated with a combined defect-light state quasi particle: the dark polariton. The complete wavefunction then is $\psi(z,t) = \cos\Theta\, E(z,t) + s(z,t)\sin\Theta$. The prefactors are defined as follows $\cos\Theta = \dfrac{\Omega}{(\Omega^2 + g^2N)^{1/2}}$ and $\sin\Theta = gN^{1/2}(\Omega^2 + g^2N)^{1/2}$ where $\Omega$ is the control field Rabi frequency, g is the defect centre-field coupling constant, N is the total number of control field photons and $s$ is the spin wavefunction, which depends on the position in the sample and on time. Adiabatic reduction of the control field Rabi frequency leads to a complete mapping of the light state on to the spins, i.e. $\psi(z,t)$ does not depend any more on E. In such an ideal scenario photon amplitude and phase would be stored in the ensemble spin state.

## Coherent control of single defect centre spin states and state read-out

One of the most intriguing aspects of the NV centre is the convenient read-out of electron spin quantum states and accessibility of their coherent control with standard ESR and NMR techniques. First, the single quantum state read-out should be discussed. Its physical basis has been already described in the introductory part of the paper. Only the $T_{0,X,Y} \rightarrow T_{1,X,Y}$ transition is visible in the fluorescence excitation spectrum. As a consequence, when the system is in the $T_{0Z}$, $T_{1Z}$ state no fluorescence can be detected. Hence each spin jump is detectable in the fluctuations of the fluorescence signal. Indeed such fluctuations have been observed (Jelezko *et al.*, 2002) and were interpreted as spin quantum jumps (see Figure 9). Since at low temperature the electron spin relaxation time is on the order of seconds, enough photons can be scattered on a particular spin quantum state to become visible. During the intermittences mostly no fluorescence (besides background) is detected. These events are marked by dtector levels below 1000 photocounts/s in Fig. 9(b). This scattering intensity indicates the spin quantum state $m_s =$

±1. The other scattering level around 15000 photocounts/s stands for the $m_s$=0 quantum state.

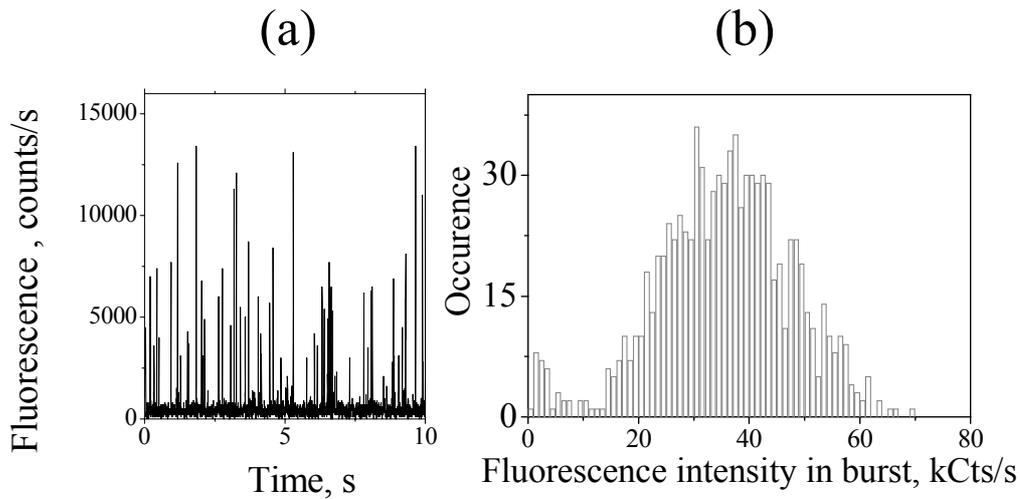

**Figure 9. Single shot spin measurements. (a) Fluorescence intensity of the NV defect under resonant excitation at T= 1.6 K. Fluorescence signal exhibits telegraph-like behaviour related to spin jumps. Note that under selective excitation centre is polarised in $m_s$=±1 ("off") state. (b) Histogram of fluorescence bursts intensities recorded after optical pumping of the defect into the $m_s$=0 state. Measurements are based on 5 ms acquisition time. Note that most of measurements outcomes correspond to the "on state" (as it expected for high fidelity readout). The small wing corresponding to "off" state is related to the finite probability of spin flip during readout process.**

The fidelity of the quantum state readout can be determined via a fluorescence counting histogram. First the $m_s$=0 was prepared using non-selective excitation. After that a short (5 ms) laser pulse was applied for state readout. For a perfect readout and state preparation, only bright ("on") state measurements are expected. Indeed experimental data indicate asymmetric fluorescence intensity distribution where "on" state is dominating. Rare events corresponding to "off" state are related to spin flip during readout (readout error). The fidelity of readout is given by the ratio of the two peaks in photon counting distribution. The amplitude of the $m_s = \pm 1$ peak is less than 5% of the $m_s = 0$ peak at the point where the two curves overlap. Hence we conclude that our readout fidelity for a given spin state is 95%. This value is comparable with those obtained for single ions (Schmidt-Kaler *et al.*, 2003).

Having established the readout technique, one might ask for the influence of the measurement on the coherence among spin states. For a single quantum system the detection of fluorescence photons must be considered as a projective measurement of the spin state. Coherence among spin states is generated by microwave irradiation in resonance with the $m_s = 0$ to $m_s = \pm 1$ transition in the electronic ground state. For reasons of experimental convenience, the investigations have been carried out at room temperature. Figure 10 shows transient nutations of a single NV electronic spin as a function of laser probe intensity. The transient (Rabi) nutations of the spin arise because of the sudden application of a microwave pulse in resonance with the ODMR transition. In a classical picture the microwave $B_1$-field generates a torque on the spin which causes the oscillations. In other words the magnetization becomes time dependent upon application of a microwave $B_1$-field amplitude according to $\vec{M} = \gamma_e (\vec{M} \, X \, \vec{B}_1)$. In a generalized interpretation of the Bloch equation it is the $M_z$ component of the magnetization, which is detected in our experiment since $\langle M_z \rangle \, \alpha \, \rho(m_s = 0) - \rho(m_s = \pm 1)$. Here $\rho(m_s = 0)$ and $\rho(m_s=\pm1)$ are the diagonal elements of the 2x2 matrix describing the spin states. The oscillations are damped out because of dephasing or an inhomogeneous distribution of transition frequencies. Upon application of a stronger laser field the damping time of the spin nutations increases significantly (see Fig. 10). This might be interpreted as a quantum Zeno effect. Let $P_{+1}= \left| m_s=\pm1 \right> < m_s=\pm1 \right|$ be the probability of $m_s=\pm1$ state occupation and $P_o = \left| m_{s=o} \right> \left< m_{s=o} \right|$ the probability of $m_s=0$ state occupation. If p is the probability to find the system in state 2 than q=1-p is the probability for measuring the system in state 1. The probability of survival in the original state is microwave pulses. Two experimental data sets recorded for different optical pumping efficiency are presented. (b) Decoherence rate as a function of optical power (circles represent experimental data. Solid line is prediction based on description of the system in terms of optical Bloch equations).

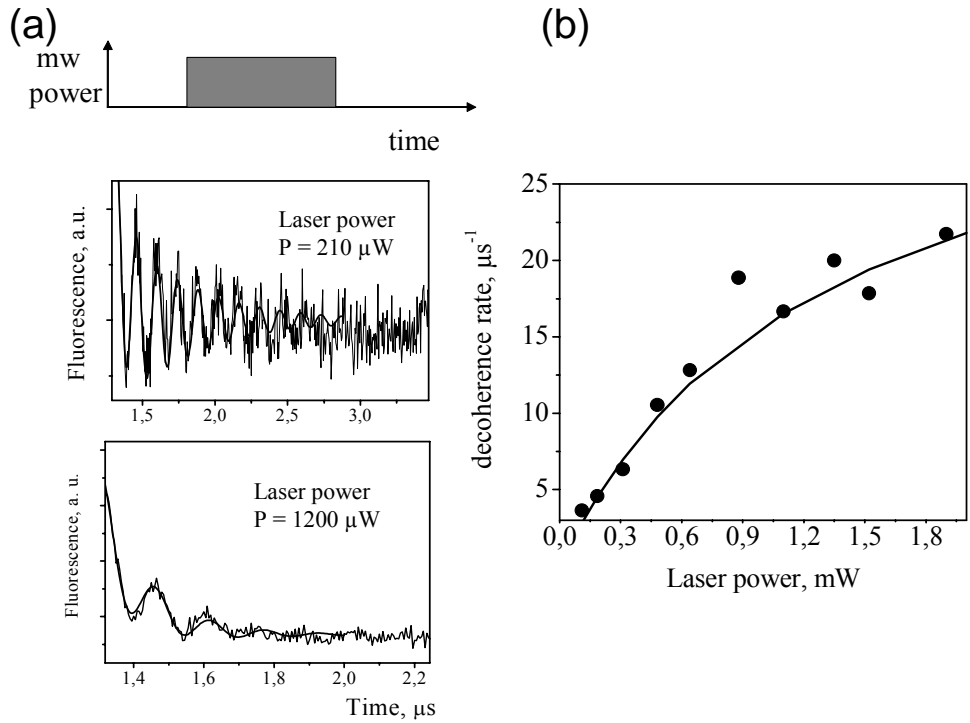

**Figure 10. (a) Transient nutations of a single electron spin associated with single NV defect under optical illumination. The upper graph shows an experimental pulse sequence. NV centre was continuously polarised using non-selective excitation. Transient nutations were induced by resonant**

$$p_{surv} = \frac{1}{2}\left(1 + (q - p)^N\right) = \frac{1}{2}\left(1 + (1 - 2p)^N\right)$$

where N is the number of measurements performed in the time interval T so that $\Delta t = \dfrac{T}{N}$ will give the time interval between two measurements. For small $\Delta t$, the probability for transition from $m_s = \pm 1$ to $m_s = 0$ is given by p and is quadratic in $\Delta t$

$$p = \lambda^2 \Delta t^2 = \left(\frac{\lambda T}{N}\right)^2.$$

For a large N, i.e. high frequency of measurements, the survival probability is therefore given by

$$p = \frac{1}{2}\left[1 + \left(1 - 2\frac{(\lambda T)^2}{N^2}\right)^N\right] = \frac{1}{2}\left(1 + e^{-2\frac{(\lambda T)^2}{N}}\right).$$

For a continuous measurement (in this case $N \rightarrow \infty$) the probability p is equal to unity. Thus a continuous measurement will inhibit the transition between two levels. As a consequence, the Rabi oscillations between two spin states will be completely suppressed as shown in Figure 10. First experimental tests of this effect have been realized experimentally by D. Wineland (Wineland *et al.*, 1995; Itano *et al.*, 1990). In the case of the NV centre, the experiments can be modelled by Bloch equations taking into account the three energy levels and the laser as well as a microwave field. In these equations the measurement process is related to detection of the laser induced fluorescence. For weak laser excitation, the decoherence rate follows a linear dependence on laser power. However for higher laser powers a saturation behaviour sets in. This is not related to the Zeno effect itself, but is due to an increased population probability of the metastable singlet state where the system gets trapped, especially at high laser intensity. In this sense the NV centre is a good model system for the Zeno effect only at low measurement frequency $(\Delta t)^{-1}$.

## Coherent spin manipulation and dephasing properties

The quantum state of the NV electron spin can manipulated with standard ESR methodology. The ground electron spin state is split by the anisotropic dipolar interaction of the two unpaired electron spins with a zero field splitting of D=2.88 GHz. The zero-field splitting tensor is of third rank and can be characterized after appropriate coordinate transformation by three main values. However, due to the dipolar nature of the interaction, the tensor is traceless such that two values are sufficient to characterize the NV centre spin ground state. If the defect would be ideally symmetric ($C_{3v}$) then the second value (usually called $E$) would be zero. In fact for quite a number of defects this holds. However, for some defects E≠0 which indicates that their symmetry obviously is lower than $C_{3v}$. In the following we will assume that $E$=0. Under this circumstance two spin sub-level *X,Y* are exactly degenerated. Linear combination of both spin sublevels give rise to either a sublevel with spin quantum number $m_s$=+1 or −1. The expectation value of $S$ in these sublevels is one. Hence it is associated with a magnetic moment and first order hyperfine coupling is present. It should be noted that this is usually not the case when no external magnetic field is applied. For the case of NV centre, the nitrogen nuclear spin couples to the electron spins of the centre resulting in a hyperfine splitting. The spin density of the ground-state electron spin wave function at the nitrogen nuclei however is low (2%) such that the hyperfine coupling to a [14]N nuclear is only around 2 MHz. This splitting and the concomitant of nuclear quadrupole splitting is only resolved in a CW ODMR experiment when low laser and microwave excitation intensities are chosen. Such well resolved spectrum is shown in Figure 11. Three lines are visible, as expected for nuclei with $I$=1 nuclear spin angular momentums. The inset in the figure shows the relevant energy level diagram together with the allowed transitions marked by arrows. Coherence among nuclear spin levels can be generated by a strong microwave pulse which excites allowed and forbidden ESR transition (branching). In a two-pulse Hahn echo experiment the interference among allowed and forbidden transition because visible as a modulation pattern. Indeed two pulse echo experiments show a characteristic modulation, which after Fourier transformation, reveals the transition frequencies observed in Figure 11. The [14]N might show a short coherence time because of its quadrupolar moment. This couples to lattice vibrations easily and hence causes phonon induced spin dephasing.

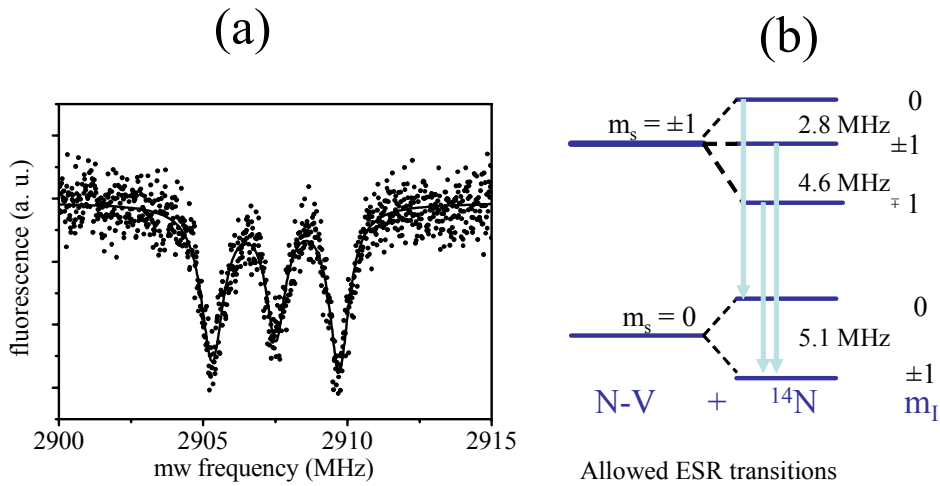

(a)

(b)

Allowed ESR transitions

**Figure 11. (a) Optically detected magnetic resonance spectra from a single NV colour centre. (b) Energy level schemes for the NV colour centre showing the hyperfine coupling in the ground state spin substructure.**

Because of hyperfine broadening of the NV centre ESR lines, it is attractive to be able to apply an as strong MW pulses as possible, i.e. achieve an as large as possible $B_1$-field. To this end either powerful MW amplifiers or miniaturized MW coils have to be used. For a one turn MW loop the achievable $B_1$ in the centre is proportional to $^1/_r$, when r is the radius of the loop. Miniaturized MW loops (r ~100 μm) were used to achieve broad band MW excitation with Rabi frequencies around 40 MHz (Jelezko *et al.*, 2004b). Going beyond this value was possible with strip line devices with gaps on the order of a few μm. With these devices, Rabi frequencies of 140 MHz have been achieved, enough to cover even hyperfine coupled $^{13}C$ spin spectra (see Figure 12). For selected samples the damping time was on the order of hundreds of μs. Hence roughly $10^4$ Rabi cycles can be observed before decoherence destroys the phase of the spin wave function.

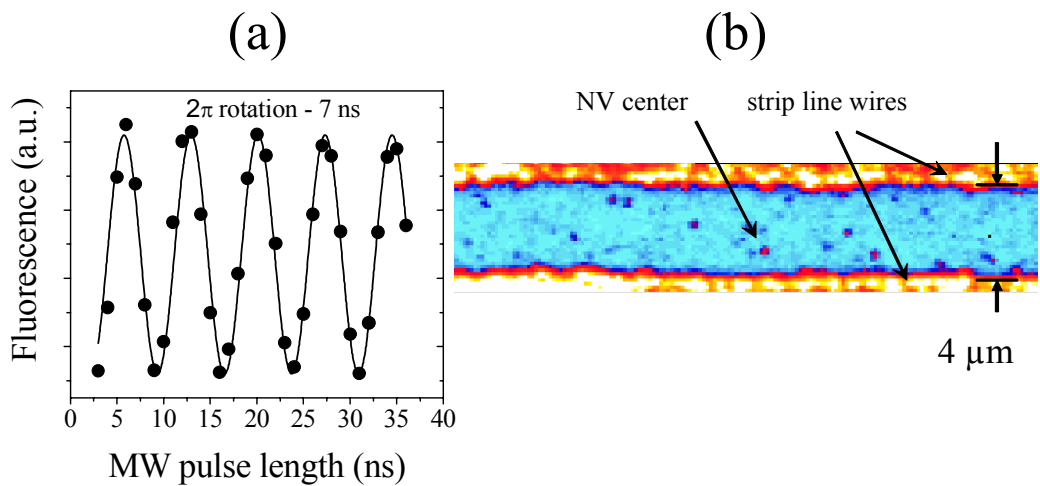

**Figure 12. (a) Rabi oscillations of a single NV defect electron spin. (b) Confocal fluorescence image of sample showing strip-line microwave wires and NV defect. Fluorescence intensity is encoded in colour scale.**

The electron spin dephasing time of the NV centre critically depends on the nitrogen concentration. A direct proportionality to the volume concentration has been found by Kennedy and co-workers (Kennedy *et al.*, 2003). Hence, experiments on NV centres with long spin phase memory require nitrogen free diamond. For this NV centres can be created by implantation of nitrogen into relatively pure type IIa diamond (Burchard *et al.*, 2005; Rabeau *et al.*, in preparation). In such samples dephasing times up to 350 µs have been found (Burchard *et al.*, in preparation).

Because of its fast decoherence and complex spin Hamiltonian the quantum state of the nitrogen nucleus is difficult to control. It is known however that the spin density at the three dangling bonds of the next nearest neighbour carbon atoms is largest. Roughly 70% of the electron spin density is expected here (Luszczek *et al.*, 2004; Pushkarchuk *et al.*, 2005). The natural abundance of the $^{13}$C I=$\frac{1}{2}$ nucleus is 1 %. Hence in a not isotopically enriched diamond it is expected that roughly one out of thirty defects should show a hyperfine coupling to a $^{13}$C nucleus. In an external $B_o$-field the spin Hamiltonian describing this system is

$$H = g_e \beta B_o + SDS + SAI + g n \beta_n B_o I.$$

Indeed, such coupling to $^{13}$C has been detected experimentally. Figure 13 shows the respective ODMR spectrum. Two EPR doublets with separation of 126 MHz are visible. The spin system needs to be described by six-level system (instead of three-level). In first order (without taking into account hyperfine coupling to nitrogen) four EPR transitions between quantum state with identical nuclear spin quantum number are allowed (see arrows in Figure 13(a)). All transitions have identical transition strength, and differences in ODMR contrast are related to frequency-selective transmission characteristics of the microwave line.

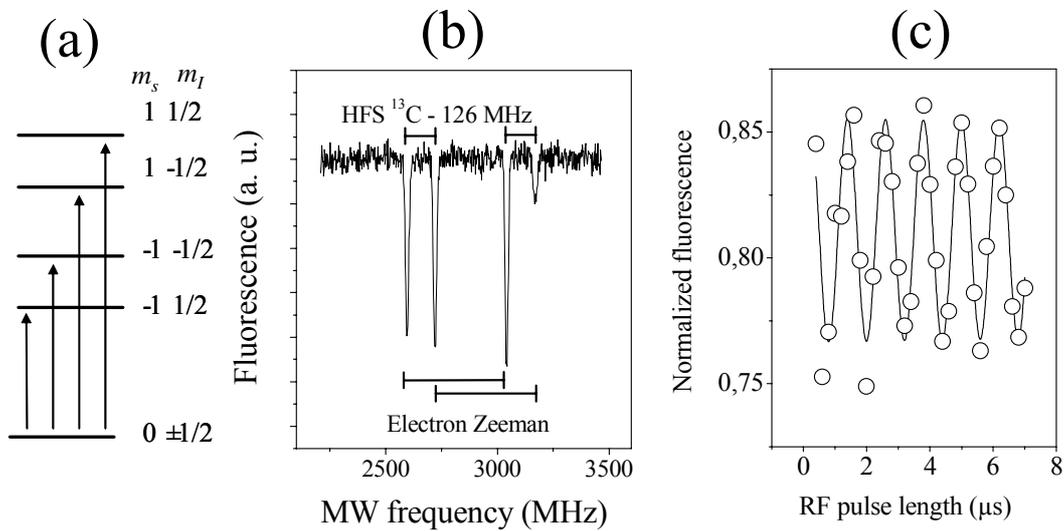

**Figure 13 (a) The ground-state energy level scheme for NV centre containing a single $^{13}$C nuclei in the first coordination shell. (b) ODMR spectrum of a single $^{13}$C coupled defect. (c) Rabi oscillations of a single $^{13}$C nuclear spin.**

Coherent electron spin nutations can be driven in this system, similar to the case shown in Fig. 12. In addition, the nuclear spin can be driven between quantum states $\left|-1, \frac{1}{2}\right\rangle$ 1 and $\left|-1, -\frac{1}{2}\right\rangle$ or $\left|1, \frac{1}{2}\right\rangle$ and $\left|1, -\frac{1}{2}\right\rangle$ respectively. Such nuclear spin quantum states do not directly couple to optical transitions because fluorescence intensity depends only on the electron spin state. Hence changes in the nuclear spin quantum states cannot be directly visualised in an ODMR experiments. However Rabi flops of nuclear spin can be monitored via the electron spin. The Rabi mutations in Fig. 13(c) are the result of such an electron nuclear double resonance (ENDOR) experiment. Due to the low magnetic field $B_o$ rather large nuclear Rabi frequencies can be observed. This effect is known in magnetic resonance as hyperfine enhancement and is based on a large electron spin contribution to the nuclear spin wave function (Schweiger and Jeschke, 2001). The important conclusion from Fig. 13 is that *both electron and nuclear spin quantum states can be manipulated coherently* and that large numbers of Rabi cycles can be observed within $T_2$. As a result either two spins quantum gates or specific electron nuclear spin quantum states can be created (Jelezko *et al.*, 2004a).

Particularly interesting is an entanglement between nuclear and electron spin. This experiment, originally performed using bulk ESR and NMR technique (Mehring *et al.*, 2004; Mehring *et al.*, 2003), can be transposed into single spin states using techniques described in following section of the paper. A specific set of two particle quantum states are the so-called Bell states

$$\phi^{\pm} = \frac{1}{\sqrt{2}} \left| \downarrow\downarrow \right\rangle \pm \left| \uparrow\uparrow \right\rangle \text{ and}$$

$$\psi^{\pm} = \frac{1}{\sqrt{2}} \left| \uparrow\downarrow \right\rangle \pm \left| \downarrow\uparrow \right\rangle .$$

A relevant energy level scheme is shown in Figure 14(a). Note that for simplicity reasons only two electronic spin states of a single $^{13}$C coupled centre are considered here. Entangled states show maximal quantum correlation among spins marked by the two arrows (first arrow indicates electron spin and the second nuclear spin state, respectively).

Although the outcome of a measurement on one spin is uncertain once that quantum state has been determined, the outcome of a measurement on the second one can be predicted with certainty.

The states $\phi^\pm$ and $\psi^\pm$ can be prepared from the electron and nuclear spin states at a single defect centre. To first order no coherent superposition between e.g. states 1 and 2 can be created with a single EPR pulse. However, electron-nuclear spin coherence can be generated in a two step process. First, electron spin coherence between states 3 and 1 is generated by a $\frac{\pi}{2}$ pulse resulting in the following state:

$$\left|\downarrow\uparrow\right\rangle \xrightarrow{\frac{\pi}{2}x} \frac{1}{\sqrt{2}}\left(\left|\downarrow\uparrow\right\rangle + \left|\uparrow\uparrow\right\rangle\right).$$

The second step now converts the electron spin coherence into electron-nuclear spin coherence. This is achieved by applying a $\pi$ pulse to the nuclear magnetic resonance transition between state1 and 2. The result of such an operation is as follows

$$\frac{1}{\sqrt{2}}\left|\downarrow\uparrow\right\rangle + \left|\uparrow\uparrow\right\rangle \rightarrow \frac{1}{\sqrt{2}}\left|\downarrow\uparrow\right\rangle - \left|\uparrow\downarrow\right\rangle.$$

This is the $\psi^-$ state described earlier. The $\psi^+$ state is created by first applying a $\frac{\pi}{2}$ pulse on the 3-4 transition and subsequently a $\pi$ pulse on the 4-2 transition. The quality of the state prepared depends on the precision of the rotation angle of individual pulses and the dephasing times. This quality needs to be checked by state tomography. For this the 4x4 density matrix is measured step by step after the initial state preparation pulse sequence.

It is assumed that at the beginning of the experiment only state 3 is populated. The difference between diagonal elements shows up as the signal strength of the respective ODMR transitions. First order coherences among states are measured after applying a $\frac{\pi}{2}$ pulse on the respective transition and subsequent measurement of the amplitude of the ODMR signal or Rabi nutation. As an example lets consider a 4x4 matrix with elements $\rho_{ij}$ which evolves in time under the action of a $\frac{\pi}{2}$ pulse. It is assumed that the coherence is present in the 1-2 transition, i.e. $\rho(t)=0$ is

$$\rho\,(t=0)= \begin{bmatrix} a & b & o & o \\ c & d & o & o \\ o & o & 1 & o \\ o & o & o & 1 \end{bmatrix}$$

Since $\rho(t)=S^{-1}\rho(t=0)S$ we find:

$$\rho(t) = \frac{1}{\sqrt{2}} \begin{bmatrix} 1 & e^{-i\phi} & o & o \\ -e^{i\phi} & 1 & o & o \\ o & o & 1 & o \\ o & o & o & 1 \end{bmatrix} \begin{bmatrix} a & b & o & o \\ c & d & o & o \\ o & o & 1 & o \\ o & o & o & 1 \end{bmatrix} \begin{bmatrix} 1 & -e^{i\phi} & o & o \\ e^{i\phi} & 1 & o & o \\ o & o & 1 & o \\ o & o & o & 1 \end{bmatrix}$$

Here a transition selective pulse on the 1-2 transition is assumed. If we just concentrate on the measurable quantity $\rho_{11}-\rho_{22}$ one gets

$$\rho_{11}-\rho_{22}=\frac{a}{2}\frac{b}{2}e^{i\phi}+ce^{-i\phi}+\frac{d}{2}-\frac{a}{2}+\frac{b}{2}e^{i\phi}+\frac{c}{2}e^{-i\phi}-\frac{d}{2}=be^{i\phi}-c_e^{-i\phi}.$$

For $\phi$, the phase angle to be $\phi = 0$ the signal measured is just $\rho_{11}$-$\rho_{22}$=b+c. For symmetric matrices the result is 2b (or 2c). Second order coherences i.e. $\rho_{2,3}$, $\rho_{3,2}$ must be converted into observable first order coherences. This is done by inverting the pulse sequence used to create the $\psi^{\pm}$ state, i.e. by using a $\pi$ pulse to swap the coherence of interest to an observable transition. This is then followed by a $\frac{\pi}{2}$ pulse on an allowed transition to obtain the same result as above, i.e. $2\rho_{23}$ or $2\rho_{32}$ for a symmetric matrix. The result of a tomography experiment on the $\psi$ state is shown in Fig. 14(b). As expected all matrix elements have about the same magnitude but the coherences show opposite sign. Fig. 14(c) depicts a calculation of the $\psi^{-}$ density matrix assuming realistic condition, i.e. finite pulse width and decoherence.

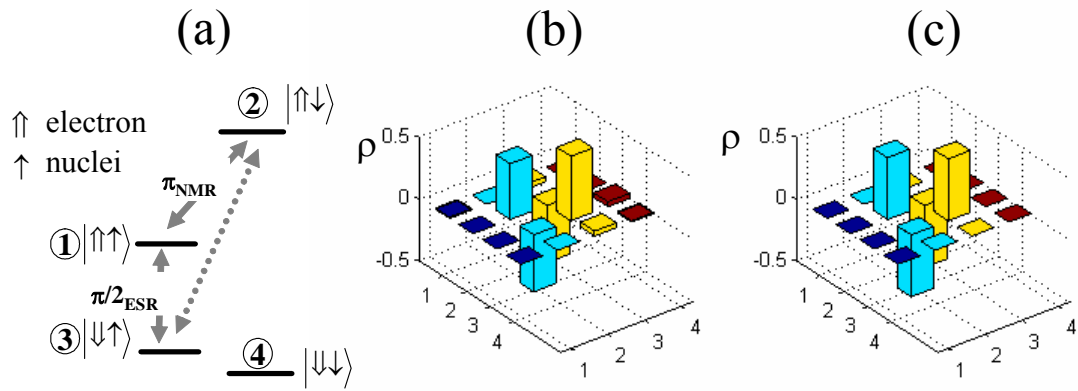

Figure 14. (a) Effective pseudospin energy level scheme and pulse sequence relevant for the generation of Bell states. The energy levels describe the interaction of a single electron with a single $^{13}$C nuclear spin in the ground state of the defect. (b) Measured density matrix of the system after preparation of $\Psi^-$ state. (c) The result of a simulation which takes into account decoherence and finite spectral width of microwave pulses.

## Conclusion

The potential of defect centres in diamond for quantum technology remains to be uncovered. Among the more than 100 luminescent defects quite a number show interesting properties for quantum information processing. As an example the Ni related defect (NE8) shows unsurpassed narrow room temperature fluorescence emission lines while preserving reasonable short excited state lifetimes and fluorescence quantum yield to ensure a large enough photon flux as single photon source. The narrow emission lines even at room temperature are a characteristic for a large number of luminescent defects in diamond which do not comprise a lattice vacancy. Due to the rigidity of the diamond lattice and the small carbon mass the Debye temperature is unusually large. As a result the phonon density even under ambient conditions is low and electron phonon coupling is small. Hence it is expected that other defects emitting at different wavelengths show similarities to the NE8. The same is true for defects with spin ground state. A number of defects other than the NV defects possess paramagnetic ground electron states. Moreover, their physical properties resemble those of the NV centres. Most notably it is expected that their spin dephasing times will be long. For the NV centre the electron spin dephasing time under ambient condition is limited by the residual paramagnetic impurities content and not by spin phonon coupling as it is often the case in other materials. The reason is very much the same as for the narrow NF8 emission: low electron- phonon coupling. Thus there is reason to believe that the currently measured value for electron spin dephasing times may be extended toward the electron spin lattice relaxation value which might be some tens of milliseconds.

Whether single electron or nuclear spin state read out is possible or not depends on the number of photons which can be scattered on a specific spin state before spin flip occurs. This in turn depends on the spin lattice relaxation rate but also on spin conservation under optical excitation. In general, readout is possible for systems with with weak spin-orbit coupling. More precisely, weak spin-orbit coupling for $S > \frac{1}{2}$ system is expected to occur in case when the magnetic dipolar coupling tensors in ground and optically excited states are collinear. Even if this is not the case, high magnetic fields might nevertheless enable single spin state readout. Hence, there is reason to believe that other than NV colour centres have potential use in quantum computing. Finally, the advances in nanotechnology will influence the impact of defects in diamond on the whole field of quantum physics in solids. Some advances have been made recently and diamond nano structuring is pursued by several groups worldwide. However, there is substantial room for improvement which makes diamond material research even more attractive than it already is.

**ACKNOWLEDGEMENTS**

The work has been supported by the European Commission (via integrated project "Quantum applications"), ARO, DFG (via SFB/TR 21 and graduate college "Magnetische Resonanz"), and the Landestiftung BW (via the programme "Atomoptik"). We acknowledge the crucial contribution of P.R. Hemmer to experiments on fast spin manipulation using strip-line structures.